\documentclass[pre,twocolumn,floatfix,showpacs,superscriptaddress]{revtex4}
\usepackage{graphicx,bm,amsmath}
\begin{document}
\title{Quantum transport and spin dynamics on shearless tori} 

\author{K. Kudo}
\email{kudo.kazue@ocha.ac.jp}
\thanks{Present address: Ochadai Academic Production, 
Ochanomizu University,
2-1-1 Ohtsuka, Bunkyo-ku, Tokyo 112-8610, Japan}
\affiliation{Department of Applied Physics, Graduate School of
Engineering, Osaka City University, Osaka 558-8585, Japan}
\affiliation{Department of Physics and Astronomy, University College
  London, Gower Street, London WC1E 6BT, United Kingdom}
\author{T. S. Monteiro}
\affiliation{Department of Physics and Astronomy, University College
  London, Gower Street, London WC1E 6BT, United Kingdom}

\date{\today}

\begin{abstract}

We investigate quantum dynamics in phase-space regions
 containing ``shearless tori''. We show that the 
properties of these peculiar classical phase-space structures
--- important to the dynamics of tokamaks --- may be exploited for
 quantum information applications. 
In particular we show that shearless tori permit the non-dispersive 
transmission of localized wavepackets. 
The quantum many-body Hamiltonian of a Heisenberg ferromagnetic 
spin chain, subjected to an oscillating magnetic field, can be reduced
to a classical 
one-body ``image'' dynamical system which is the well-studied Harper map.
The Harper map belongs to
 a class of Hamiltonian systems (non-twist maps) which contain shearless
 tori. We show that a variant with sinusoidal 
time driving ``driven Harper model'' produces shearless tori which are
especially suitable for quantum state transfer. The behavior of the
 concurrence is investigated as an example.  

\end{abstract}

\pacs{75.10.Pq, 03.67.Mn, 05.45.Mt}

\maketitle

For a Hamiltonian system, the onset of classical chaotic dynamics is
associated with the disappearance of phase-space barriers termed
invariant tori: when the last invariant torus disappears, the  chaotic
diffusive motion is unbounded (``global''). 
For a wide class of classical dynamical systems, described by so-called
``twist-maps'' and exemplified by the all-important Standard Map
\cite{Ott}, there is a single threshold for this process, i.e., once the
last invariant torus breaks, the dynamics is globally diffusive for all
parameters above the threshold. 
However, for another class of dynamical systems described by ``non-twist
maps'', this is not generally the case. 
A new class of tori, termed ``shearless tori'' \cite{Shinohara} can be
present.  The { \em classical} properties of these  shearless tori are
attracting considerable attention, due in part to their possible
relevance to improved confinement of fusion plasmas in  tokamaks
\cite{tokamaks}, but also because the properties of non-twist maps are
less well studied. 
New dynamical phenomena such as separatrix reconnection lead to the
intermittent re-appearance of the shearless tori which, when present,
separate different regions of the chaotic phase space.

In a quite different context, studies of quantum dynamics of spin
chains, for example of quantum state transfer and entanglement
generation, now play a central role in the burgeoning field of quantum
information \cite{Bose}.  
There is also growing interest in  potential applications of nonlinear
dynamics  in quantum information. 
In particular, the realization that a many-body Hamiltonian can be
analyzed with the dynamics of  ``one-body image'' quantum and 
classical Hamiltonians \cite{Prosen,Boness} has proved very insightful. 
For example, the Heisenberg ferromagnetic spin chain, in a pulsed
parabolic magnetic field, has the Standard Map as its classical image
\cite{Boness}, provided {\em position} coordinates in the many-body
Hamiltonian are mapped onto {\em momenta} of the image system and
vice-versa. If, instead,  a pulsed sinusoidal external magnetic field is
applied and the canonical coordinates are not interchanged, the spin
system maps rather onto one of the best-known  non-twist maps: 
the kicked Harper map. The quantum dynamics of the kicked Harper map
have already formed the subject of several studies \cite{Harper,Dana}. 
From the viewpoint of entanglement screening, recently, calculation of
the concurrence showed that nonlinear resonances in a classically mixed
phase space induce robust entangled states \cite{Buch}.  

However, to date, the quantum dynamics near shearless tori,
and their potential for quantum information applications have never been
investigated.  
In the present study, we have calculated quantum transport along
shearless tori and found it to be quite different from the dynamics
along ordinary tori. 
In particular, we find that shearless tori provide non-dispersive
transport, while ordinary tori are dispersive. In other words, for a
spin chain, a spin wavepacket initially localized with starting
conditions close to an ordinary torus, will rapidly become completely
delocalized along the whole length of the spin chain. 
In contrast, for starting conditions near a shearless torus,
an initially localized Gaussian wavepacket largely preserves its shape
for long times, but moves along the spin chain at a near constant rate.  
In the chaotic regime, the shearless tori provide isolated regular
`channels' through the chaotic sea which can be used to propagate
coherent spin states (approximately) non-dispersively.  
We show that, for a driven Harper model, one may even start the
wavepacket with zero momentum. We attribute the non-spreading of  
quantum wavepackets near shearless tori to a local quasienergy spectrum
(i.e., for the Floquet states  localized on the shearless tori) which is a
perturbed harmonic spectrum. 
The corresponding evolution of the concurrence was also obtained.

We consider the Heisenberg ferromagnetic spin chain under an oscillating
magnetic field.
Here, we concentrate on a spin-1/2 and one-down-spin case, where a state
is in a superposition of the states where one spin is down and the others
are up.
In that case, the model corresponds to a
tight-binding model with an oscillating field of amplitude
$B_0$ by the Jordan-Wigner transformation. 
Since the total $S_z$ is conserved, i.e. the one-down-spin state is
conserved, the correlation terms disappear. 
Then, the Hamiltonian is written as
\begin{equation}
H(t) = \frac{J}{2}\sum_{j=1}^N
(c_j^{\dagger}c_{j+1}+c_{j+1}^{\dagger}c_j)
 + B_0 F(t)  \sum_{j=1}^N\cos\left( \frac{2\pi}{N}j \right)
c_j^{\dagger}c_j.
\label{eq:tight}
\end{equation}
Here, $c_j^{\dagger}$ and $c_j$ are the creation and annihilation
operators of a fermion at the $j$th site, respectively. 
For ferromagnetic cases, $J<0$.
The total number of sites is $N$, and  periodic boundary conditions are
imposed. $F(t)$ is a periodic function of time with period $T$. 

\begin{figure}
\includegraphics[width=5cm]{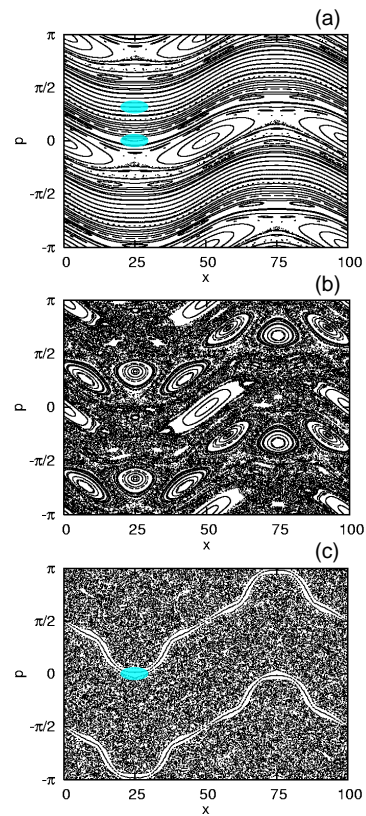}
\caption{\label{Fig1} (Color online) 
Classical phase space for (a) $\omega=0.20$ (near
 integrable), (b) $\omega=0.16$ mixed phase-space with no apparent tori,
 and (c)  $\omega=0.12$ showing the re-appearance of shearless tori:
 phase space is almost entirely chaotic but regular channels,
 containing shearless tori, which separate different chaotic regions,
 are seen.
The colored patched areas correspond to the initial wave packet of
 quantum dynamics in Fig.~\ref{Fig2}.} 
\end{figure}

We solve spin dynamics from Eq.~(\ref{eq:tight}). However, to analyze
the quantum behavior, we first note this:
In the absence of the time-dependent external field, the model
(\ref{eq:tight}) is associated with a dispersion relation, obtained from
the Bethe ansatz, $E_i=J(1- \cos k_i)$  with $i=0,1,2,\ldots,N$ and
where the $E_i$ are energy eigenvalues and the $k_i$ quantized
wavenumbers. Equating the $k_i$ with the momenta of a classical system
\cite{Prosen,Boness}, i.e., $k_i \to p$ and taking a continuous position
variable $x_j=\frac{2\pi}{N}j \to \frac{2\pi}{N} x$, we can see that
Eq.~(\ref{eq:tight}) maps onto the one-body classical ``image''
Hamiltonian:  
\begin{equation}
H(x,p,t)=J\cos p +B_0F(t) \cos [(2\pi/N)x].
\label{eq:harper}
\end{equation}
If $F(t) = \sum_n \delta(t-nT)$, we have the usual kicked Harper model. 
Here, we take $F(t) = \sin \omega t$, with $\omega=2\pi/T$, and
investigate a modification we term the ``driven-Harper model''.
The corresponding classical equations of motion are given by:
\begin{equation}
\left\{
   \begin{array}{lc}
     \dot{x}=-J\sin p & \quad (0< x \le N),\\
     \dot{p}=(2\pi/N)B_0\sin\omega t\sin[(2\pi/N)x]
             &  \quad (-\pi< p \le \pi ).
   \end{array}
\right. 
\label{eq:cl}
\end{equation}
Integrating Eq.~(\ref{eq:cl}) numerically, we obtain 
$x(t)$ and $p(t)$. Classical surfaces of section (SOS) are 
obtained by plotting values of $(x,p)$ at
 integer multiples of the period, $t=n(2\pi/\omega)$, where $n$ is
an integer. Three SOS are shown in Fig.~\ref{Fig1}. 
Here, we set $J=-1$, $B_0=2$, $N=100$, and these values are fixed
below. With decreasing 
frequency $\omega=0.20$, $0.16$, $0.12$, the classical phase space becomes
increasingly chaotic; at $\omega=0.16$ all tori have disappeared;
however, at $\omega=0.12$, shearless tori have re-appeared, embedded 
in two regular `channels' running through a fully chaotic 
phase space (bar a few very small stable islands). 

For the kicked Harper map, shearless tori occur for 
$p \approx \pm \pi/2$ \cite{Shinohara}. 
They correspond to an extremum of the rotation number; here 
$\frac{{\partial {\dot x}}}{{\partial p}}=0$ at $p \simeq \pm \pi$.
The shearless tori for $\omega=0.12$ are extremely distorted and
span the full range of the map, from $p= 0$ to $\pm\pi/2$. The kicked
Harper map, on the other hand, does not produce shearless tori which
include the $p=0$ region, so for this reason, using a driven rather than
a kicked spin chain is advantageous. In other words, to prepare a
spin-wave packet on the shearless torus region, one should excite low
energy $E(k_j) \simeq 0$ spin-waves (i.e., $k_j\simeq 0$ since 
this corresponds  
to the bottom of the $E(k_j)$ energy band) in the region of the spin chain
containing a node of the magnetic field (i.e., 
$x_j \approx \pi/2$ or $x_j \approx 3\pi/2$).
Exciting low-energy spin waves is a simpler procedure than selecting
a more precise range of higher-energies, which would be required if the 
shearless tori are only found at $p \approx \pi/2$ as is the case for
the kicked system. 
 
Returning now to our spin-Hamiltonian, Eq.~(\ref{eq:tight}), we compare
the time evolution of a spin wavepacket with initial conditions
corresponding, in the image classical system, to shearless tori with
those of normal tori. We consider the quantum spin distribution:  
\begin{equation}
P(j,t)=|\langle j|\psi (t)\rangle |^2.
\end{equation}
where $|j\rangle$ is the state where the $j$th spin is down, and
$|\psi(t)\rangle$ is calculated from Eq.~(\ref{eq:tight}).
Our initial state $|\psi(t=0)\rangle$ is a Gaussian spin wavepacket
with the phase space area corresponding to the classical plot in
Fig.~\ref{Fig1}. 
The center of the wavepacket is denoted by $j_0$ and $k_0$ in the real and
momentum space, respectively. The width of the wavepacket is set to be
$\Delta_j=5$. 

\begin{figure}
\includegraphics[width=6cm]{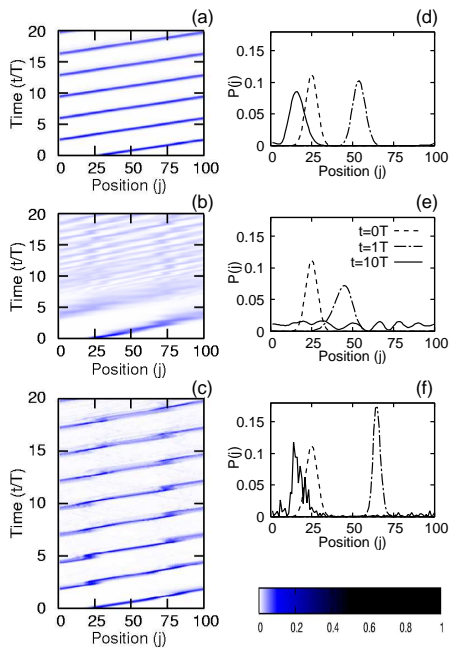}
\caption{\label{Fig2} (Color online) 
Time evolution of a quantum spin-wavepacket, initially centered at $j_0=25$,
showing dispersive propagation on normal tori and non-dispersive
 propagation on shearless tori. 
(a) $\omega=0.20$ and $k_0=1.0$; wavepacket prepared
 initially on shearless torus of Fig.~\ref{Fig1}(a), but in the near
 integrable regime.  
(b) $\omega=0.20$ and $k_0=0.0$; initial state is on an ordinary torus
 of Fig.~\ref{Fig1}(a)   
 and rapidly delocalizes along the length of the spin-chain. 
(c) $\omega=0.12$ and $k_0=0.0$; the spin wavepacket initially on
 shearless torus embedded in the chaotic sea of Fig.~\ref{Fig1}(c) 
 propagates largely without dispersion, but the effect is less pronounced
 than in the integrable case.
(d), (e), and (f) correspond to (a), (b), and (c), respectively. 
They exhibit the form of wavepackets after $t=0T$, $1T$, and $10T$.}
\end{figure}

In Fig.~\ref{Fig2}(a), we show the behavior (for $\omega=0.20$) of a
wavepacket initially centered on $j_0=25$ and $k_0=1.0$ (i.e., 
mapping onto a regular phase-space region in the image classical system
including a shearless torus). the distribution shows very little
delocalization for as long as the numerics were run (20 periods).
Despite the same frequency, the distribution in Fig.~\ref{Fig2}(b),
which initially centered on the region corresponding to normal tori in
the classical map, rapidly delocalizes and there is no discernible
localization at $t=10T$. However, Fig.~\ref{Fig2}(c) (for $\omega=0.12$)
shows again little delocalization of the wavepacket initially centered
on $j_0=25$ and $k_0=0.0$ (i.e., on a shearless torus). 
Numerical experiments show that an equivalent 
classical ensemble of particles around a shearless torus spread slowly: 
the lack of shear
also produces a strong classical effect.

One can attempt to understand the quantum behavior from its  
quasi-energy spectrum: as this is a temporally 
periodic system, it is the eigenstates of the one-period unitary
evolution operator (Floquet states) $\phi_m(x,t)$ and associated
eigenphases (quasi-energies) $\varepsilon_m$  which represent the
stationary states of the system.
One should first emphasize that the non-dispersive behavior is not
simply a wavepacket revival effect \cite{revival} where a wavepacket may
partially or totally regain its shape after some revival time, 
dependent on the underlying frequencies. In our case, the wavepacket
does not ever lose its essential shape. 
The behavior is rather analogous to motion of a wavepacket on a harmonic
potential: the wavepacket oscillates in position without spreading; 
if the shape of the underlying potential is different (so our $\Delta_j$
is not the width of its ground state) the wavepacket exhibits only a
certain  ``breathing'' in its width, demonstrated in \cite{revival}. 
The modest variability in the wavepacket shape seen in Fig.~\ref{Fig2}
would be compatible with such breathing. 

\begin{figure}
\includegraphics[width=4cm]{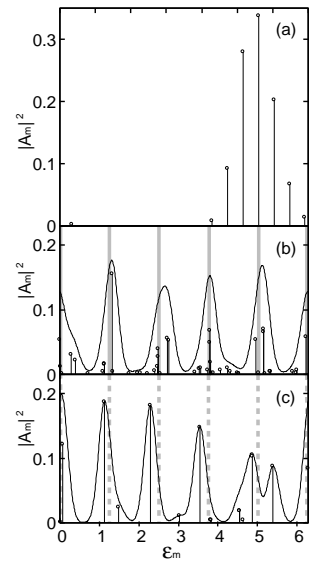}
\caption{\label{Fig3} Quasi-energy spectrum local to shearless
tori, showing that it corresponds to a near harmonic spectrum in
the near-integrable limit and a perturbed harmonic spectrum in
the chaotic limit. The spectrum is plotted for $|A_m|^2>10^{-3}$.
The low resolution spectrum is a Gaussian-smoothed one.
The gray broad bars are just an eye guide, and their interval is $2\pi/5$.
(a) Quasienergy spectrum for $\omega=1.0$ showing
equal quasi-energy spacing. 
(b) Low resolution (smoothed) spectrum
near shearless tori embedded in chaotic sea for $\omega=0.12$
showing the spectrum
approximates a perturbed harmonic spectrum. 
(c) Low resolution spectrum near ordinary torus at $\omega=0.20$.}
\end{figure}

We have calculated the local Floquet spectrum for the initial Gaussian
wavepacket, i.e., the overlap between the wavepacket and the Floquet
states, $A_m = \langle\phi_m(x,t=0)| \psi_0\rangle$. The probability
$|A_m|^2$ is plotted in Fig.~\ref{Fig3} as a function of
quasi-energy. For the wavepacket on a shearless torus in a
near-integrable regime, shown in Fig.~\ref{Fig3}(a),  a `harmonic'
ladder of equally-spaced eigenstates is evident. In
Fig.~\ref{Fig3}(b), the shearless torus embedded in the chaotic regime,
the low-resolution smoothed spectrum shows a set of equally
spaced peaks. 
The interval of the peaks is $2\pi/5$. This corresponds to the fact that
the wave packet returns almost the same position as the initial one
after 5 cycles ($t=5T$). 
We propose that this represents a perturbed harmonic spectrum found for
eigenstates local to shearless tori, in the near-integrable regime. In
the chaotic regime [Fig.~\ref{Fig3}(b)], 
the eigenstates are no longer completely localized
on the shearless torus region and overlap with the chaotic regions;
conversely, eigenstates with most support from the chaotic regions
overlap with the torus region.  
Consequently, the states no longer form a ladder of individual states, 
but their amplitude is split redistributed into a cluster of nearby
states. Nevertheless, the presence of these regularly spaced clusters is
sufficient to hinder dispersion of the wavepacket for a considerable 
time.
For the ordinary torus, however,
the spacing is irregular and even with smoothing, and the low-resolution
spectrum has no equal spacing as shown in Fig.~\ref{Fig3}(c). 

\begin{figure}
\includegraphics[width=8cm]{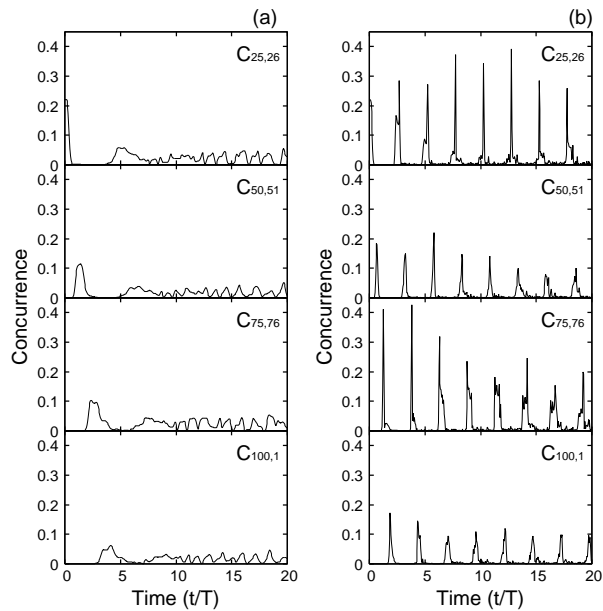}
\caption{\label{Fig4} Time evolution of the concurrence, contrasting
entanglement transport by wavepackets moving along normal tori 
(left panels) with 
the non-dispersive shearless tori (right panels).  
(a) $\omega=0.20$ showing broad small peaks
corresponding to Fig.~\ref{Fig2}(b),
and (b) $\omega=0.12$ showing periodic sharp peaks
corresponding to Fig.~\ref{Fig2}(c). }
\end{figure}

The entanglement is also transported when the Gaussian packet travels.
We employ the concurrence $C_{i,j}$, which is a measure of the bipartite
entanglement of two sites $i$ and $j$~\cite{wootters}.
We concentrate on the cases corresponding to Figs.~\ref{Fig2}(b) and
\ref{Fig2}(c) as we are interested in the
transport of quantum information. 
Figure~\ref{Fig4} shows the time evolution of the concurrence,
$C_{25,26}$, $C_{50,51}$, $C_{75,76}$, and $C_{100,1}$ for (a)
$\omega=0.20$ and (b) $\omega=0.12$. Here, we should notice that the
100th and 1st sites are neighboring sites because of the periodic
boundary conditions. 
In Fig.~\ref{Fig4}(a), no sharp peaks appear, although we can
see some small peaks. 
On the other hand, in Fig.~\ref{Fig4}(b), the concurrence has sharp
periodic peaks. Moreover, several peaks are much higher than the initial
value, $C_{25,26}(t=0)\simeq 0.22$. 
This behavior of the concurrence corresponds closely to that seen in the
spin dynamics. 

In conclusion, we find that quantum transport using shearless tori
has a very different character from that associated with normal tori.
The quantum spin chain under an oscillating field can,
for certain conditions, provide approximately non-dispersive
transport even in a largely chaotic regime. 
The non-dispersion of Gaussian states is a property associated with
large stable islands, in fact. These have been shown recently to 
preserve entanglement against noise \cite{Buch}. The shearless tori
may well provide a similar advantage while also transporting 
entanglement around the spin chain.

The authors thank K. Nakamura and S. Bose for
helpful discussions and K.K. thanks the JSPS for support.

\end{document}